\documentclass{emulateapj}
\usepackage{lscape}
\usepackage{rotate}

\shorttitle{Clusters in the Zone of Avoidance}
\shortauthors{Kocevski, Ebeling, Mullis, \& Tully}

\begin{document}

\title{Mapping Large-Scale Structures Behind the Galactic Plane:  \\ The Second CIZA Subsample}
\author{Dale D. Kocevski, Harald Ebeling, Chris R. Mullis\altaffilmark{1}, \& R. Brent Tully}

\affil{Institute for Astronomy, University of Hawaii, 2680 Woodlawn Dr., Honolulu, HI 96822}
\altaffiltext{1}{Department of Astronomy, University of Michigan, 501 E. University Ave., Ann Arbor, MI 48109}
\email{kocevski@ifa.hawaii.edu; ebeling@ifa.hawaii.edu; cmullis@umich.edu; tully@ifa.hawaii.edu}

\begin{abstract}

We present the latest results of the Clusters in the Zone of Avoidance (CIZA) survey, which is mapping the large-scale matter distribution behind the Milky Way by performing the first systematic search for X-ray luminous galaxy clusters at low Galactic latitudes.  The survey's approach, which uses X-ray emission to locate cluster candidates, overcomes the problems faced by optically selected cluster surveys which have traditionally avoided this region of the sky due to the severe extinction present along the Galactic plane.  We here present the second flux-limited CIZA cluster catalog containing 60 X-ray luminous galaxy clusters, 88\% of which are new discoveries.  We also examine the degree to which known superclusters extend into the Zone of Avoidance and highlight newly discovered structures which have previously gone unnoticed.  We show that the survey has found far fewer rich clusters in the Great Attractor region than would be expected given the region's proposed mass.  Instead, we find a significant increase in the number of clusters \emph{behind} the Great Attractor, with the most notable being an association of clusters near the Shapley supercluster.  We propose these clusters trace an extension of the large-scale filament network in which the Shapley concentration is embedded.  We also highlight an association of clusters near the Galactic anticenter, which is the first supercluster found to be completely hidden by the Milky Way.    Our finding of a less massive Great Attractor and the detection of significant structures behind the complex supports studies which suggest the motion of nearby galaxies, including that of the Local Group, is due, in part, to a large-scale bulk flow which is induced by overdensities beyond the Great Attractor region.

\end{abstract}

\keywords{galaxies: clusters: general --- large-scale structure of universe --- X-rays: galaxies: clusters}

\section{Introduction}

Galaxy clusters are the nodes of the cosmic web which permeates the Universe, making them probes of the organization of matter on the grandest scales.  While galaxies have been the preferred tracer of the local distribution of matter (Geller \& Huchra 1989; Saunders et al. 1995; Colless et al. 2001; York et al. 2000), clusters offer a complementary means to trace the large-scale structure of the Universe to distances beyond the effective range of galaxy redshift surveys (Tully et al. 1992; Mullis et al. 2001).  Although clusters more sparsely sample the cosmic web, the similarity in the shape of the galaxy and cluster power spectrums suggest both are effective in tracing the same underlying density field (Peacock \& Dodds 1994; Tadros et al. 1998).  The advantage offered by clusters is that their significant X-ray luminosities and galaxy overdensities allow for the compilation of volume-limited samples out to extremely large distances ($r > 200 h^{-1}$ Mpc).  In addition to the study of large-scale cosmography, such samples provide a means to investigate a range of cosmological issues.  The location of clusters provides insight into the spatial distribution of the initial fluctuations which are thought to have seeded the current cluster population in early epochs, the mass function of clusters offers details on the amplitude spectrum of those primordial fluctuations, and the evolution of the number density of clusters with redshift can constrain fundamental cosmological parameters.  More locally, the deep gravitational wells of clusters act as accelerators of large-scale flows and knowledge of their distribution provides a means to reconstruct the peculiar velocity field (Branchini \& Plionis 1996; Kocevski et al. 2004, 2005).

The largest cluster sample used for many such studies has been the optically selected Abell/ACO catalog (Abell 1958; Abell, Corwin \& Olowin 1989, hereafter ACO).  The Abell/ACO sample has been widely employed due to its significant sky coverage (2/3), large characteristic depth ($\sim240 h^{-1}$ Mpc) and relatively robust cluster identification criteria compared to other visually compiled samples such as the Zwicky catalog (Zwicky et al. 1961-68).  More recently, moderately deep X-ray cluster surveys over large parts of the sky have been carried out using data from the \emph{ROSAT} All-Sky Survey (RASS, Tr\"{u}mper 1993), which is the first and only survey to provide X-ray imaging data over the entire sky.  While an array of cluster samples have been compiled from the RASS (Romer et al. 1994; Pierre et al. 1994; Burns et al. 1996; Ebeling et al. 1996, 1998, 2000, 2001, 2002; Scharf et al. 1997; De Grandi 1999; Henry et al. 1997; Vikhlinin et al. 1998; Ledlow et al. 1999; B\"{o}hringer et al. 2000; Mullis et al. 2003, 2004), the largest and most complete are currently the extended Brightest Cluster Sample (eBCS, Ebeling et al. 1998, 2000) and the \emph{ROSAT}-ESO Flux Limited X-ray catalog (REFLEX, B\"{o}hringer et al. 2004). Covering the northern and southern hemispheres, respectively, the eBCS and REFLEX samples contain a combined 737 clusters with X-ray fluxes above $3\times 10^{-12}$ erg cm$^{-2}$ s$^{-1}$ and redshifts of $z\leq 0.3$.

\subsection{The Zone of Avoidance}

Since finding overdensities in the galaxy distribution becomes increasingly difficult at low Galactic latitudes due to the severe extinction and stellar obscuration present in the direction of the Milky Way (MW), optically selected catalogs such as Abell/ACO are plagued with poor coverage in a $40^{\circ}$ wide strip centered on the plane of the Galaxy known as the Zone of Avoidance (ZOA).  Unfortunately, despite the decreased extinction through the MW at X-ray wavelengths, both the eBCS and REFLEX surveys have followed the precedent set by optically selected surveys and have also avoided the Galactic plane.  With the eBCS and REFLEX surveys limited to Galactic latitudes of $|b|>20^{\circ}$, the Galactic plane, which covers 1/3 of the sky, is the final region in which the cluster distribution remains to be mapped.  This is particularly troubling since there is considerable dynamical evidence that the ZOA harbors some of the most massive large-scale structures in the local Universe.  Both the Local Group's (LG) peculiar velocity (Kogut et al. 1993) and the motion of galaxies within $40 h^{-1}$ Mpc of the MW (Lynden-Bell et al. 1988, hereafter LB88) suggest galaxies in the local volume are flowing toward a vertex behind the Galactic plane, presumably due to the gravitational influence of overdensities in this region.  Assuming this motion was due to infall into a single ``Great Attractor'' (GA), LB88 estimated that the source of the flow was located roughly $43 h^{-1}$ Mpc away and near the Hydra-Centaurus cluster association.  This distance, coupled with the large peculiar velocities observed in nearby galaxies, implied the rather high mass of $\sim5\times10^{16} h^{-1}_{50}$ M$_{\odot}$ for the GA complex.  LB88 found that such a relatively nearby, massive overdensity would be responsible for $\sim70\%$ of the LG's observed peculiar velocity.

The LB88 findings have been controversial since subsequent redshift surveys that have encompassed the GA region have failed to detect a mass overdensity as large as the one implied by the LB88 peculiar velocity data (Dressler 1988; Strauss et al. 1992; Hudson 1993, 1994), nor have they conclusively measured the backside infall into the GA one would expect if the region were best described as a single, stationary attractor (Mathewson et al. 1992, Courteau et al. 1993).  More recent work has suggested that a significant component of the LG's peculiar velocity is in the form of a large-scale bulk flow which continues past the GA region and is induced by even more distant structures in the ZOA (Zaroubi et al. 1999; Tonry et al. 2000; e.g. Hudson et al. 2003).  

The difficulty in determining the source of the LG's motion and the possible bulk flow in which it participates is largely due to the incompleteness present in the direction of the MW.  With cluster surveys avoiding the Galactic plane, the distribution of large-scale structures which are the accelerators of such flows in this region has remained largely unknown.   A variety of techniques have been used to reconstruct the ZOA, ranging from uniform filling (Strauss \& Davis 1988; Lahav 1987) to a spherical-harmonics approach which extends structures above and below the plane into the ZOA (Plionis \& Valdarnini 1991, cf. Brunozzi et al. 1995), but the value of these reconstruction techniques is limited if the MW does indeed obscure dynamically significant structures, as has been suggested.

\begin{figure*}[t]
\epsscale{1.1}
\plotone{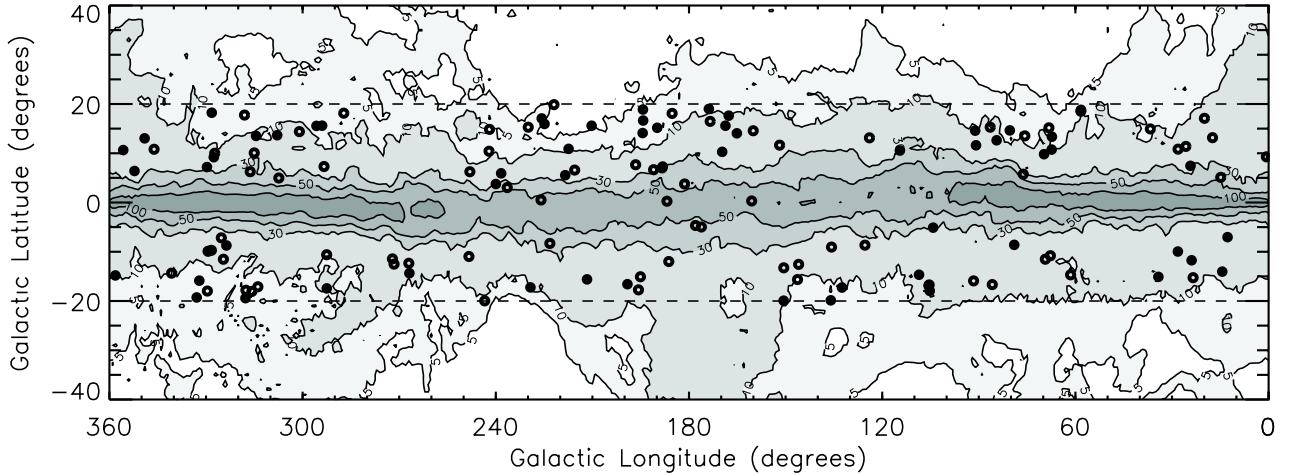}
\caption{Distribution of the current CIZA sample along the Galactic plane overlayed on contours of Galactic neutral Hydrogen column density in units of $10^{20}$ cm$^{-2}$.  Open circles denote the location of B1 clusters, while filled circles represent B2 clusters.  The current CIZA sample contains 133 clusters with X-ray detect fluxes greater than $3\times 10^{-12}$ erg cm$^{-2}$ s$^{-1}$.} 
\end{figure*}

In order to reveal the nature of regions such as the GA and other large-scale structures hidden behind the MW, we began the Clusters in the Zone of Avoidance (CIZA) project to map the cluster distribution behind the Galactic plane for the first time.  Our approach, which uses X-ray emission to locate cluster candidates, overcomes the problems faced by optically selected catalogs at low Galactic latitudes due to the decreased extinction at X-ray wavelengths.  The strategy of the CIZA survey is consistent with that of the eBCS and REFLEX surveys and therefore the completion of CIZA project will facilitate the construction of the first all-sky, X-ray selected cluster sample.  In this paper we present the latest results from the CIZA survey.  This is the second in a series of papers regarding the CIZA project; Ebeling, Mullis \& Tully (2002, hereafter Paper I) recently introduced the survey and presented a subsample of the 73 brightest CIZA clusters (the B1 sample).  We have since extended the survey to lower fluxes and present here the next CIZA subsample (the B2 sample) as well as a progress report on the survey, which is nearing completion.  In what follows, we describe the CIZA survey in more detail in \S2, present the B2 sample in \S3, and discuss the large-scale cosmography of the cluster distribution behind the Galactic plane in \S4.  Finally the future of the CIZA survey and our concluding remarks are presented in \S5.  Throughout this paper we use $H_{0}=50$ $h$ km s$^{-1}$Mpc$^{-1}$ for the calculation of comoving distances and all listed X-ray fluxes are given in the 0.1--2.4 keV \emph{ROSAT} band unless otherwise noted.

\section{The CIZA Survey}

The CIZA survey is the first systematic search for X-ray luminous galaxy clusters behind the Galactic plane.  The project uses X-ray data from the RASS for initial target selection and subsequent optical and near-infrared observations to confirm or refute the cluster nature of all selected candidates.  This approach has proven highly effective in narrowing the gap in the cluster distribution in the direction of the MW.  Thus far the CIZA survey has found and spectroscopically confirmed 205 galaxy clusters in the Galactic plane.  In this section we describe the strategy and overall progress of the CIZA survey.




\subsection{Survey Strategy}

The CIZA survey strategy is discussed in detail in Paper I and therefore we only briefly summarize it here. Initial target selection for the survey is based on X-ray detections in RASS Bright Source Catalog (BSC, Voges et al.\ 1999).  Candidate clusters are selected from the BSC if they (1) are located in the ZOA ($|b|<20^{\circ}$), (2) have a nominal ``detect'' X-ray flux in excess of $1\times 10^{-12}$ erg cm$^{-2}$ s$^{-1}$, and (3) fall above a spectral hardness ratio cut, which is meant to discriminate against softer, non-cluster X-ray sources\footnote{See Figure 1 of Paper I}.  

Applying these three selection criteria we extract
a preliminary sample of sources from the BSC, which are then cross-correlated against existing databases to identify sources with
known clusters as well as those with cataloged non-cluster counterparts.
All remaining unidentified sources (both apparent clusters and
sources of ambiguous nature) are the subject of a comprehensive
imaging survey, primarily carried out in the optical R-band.
All clusters confirmed in this imaging survey are then targeted in spectroscopic observations to
measure redshifts for at least two cluster members.  The majority of
these optical follow-up observations are carried out with the
University of Hawaii 2.2m telescope.  Cluster candidates in the 20\% of
the survey area that are not observable from Mauna Kea have been targeted
with the Anglo-Australian 3.9m, the CTIO 1.5m and 4.0m, the ESO NTT 3.5m, 
and the SOAR 4.0m telescopes.

The initial ``detect'' flux of candidate clusters is based
on the total net count rate quoted in the BSC as detected within a
5 arcmin radius circular aperture. We convert these detect count rates to approximate detect fluxes
assuming a generic cluster spectrum, namely the spectrum of a thermal
plasma with ${\rm k}T=4$ keV, a metallicity of 0.3 solar, and the
appropriate equivalent nuetral Hydrogen column density, $n_{\rm H}$, in the direction of the
source from Dickey \& Lockman (1990). We refer to these count rates
and fluxes as the ``BSC detect'' values; total fluxes based on the
total net count rate within a metric aperture of 1.5 $h_{\rm 50}^{-1}$
Mpc (radius) are computed for confirmed clusters once redshift information 
becomes available (see \S3.2).



\subsection{Survey Progress}

Thus far the CIZA project has spectroscopically confirmed 205 galaxy clusters in what used to be the ZOA.   The recently published B1 sample contains the 73 X-ray brightest of these clusters which have detect fluxes greater than $5\times 10^{-12}$ erg cm$^{-2}$ s$^{-1}$.  The B2 sample presented in this paper contains 60 clusters with detect fluxes between 3 and $5\times 10^{-12}$ erg cm$^{-2}$ s$^{-1}$, while the remaining 72 clusters fall into the final subsample ($f_{\rm X, BSC} < 3\times 10^{-12}$ erg cm$^{-2}$ s$^{-1}$) which is nearing completion.  While the survey has been successful in redetecting all previously known clusters in the ZOA, such as the Ophiuchus, Abell 3526, Cygnus-A, and Triangulum Australis clusters, $79\%$ of the clusters found thus far have been new discoveries.   

The location of the B1 and B2 clusters on the plane of the Galaxy is shown in Figure 1, along with the distribution of Galactic $n_{\rm H}$ from Dickey \& Lockman (1990).  We find that the CIZA survey is quite effective in finding clusters in regions of moderate extinction ($n_{\rm H} < 50\times 10^{20}$ cm$^{-2}$), which make up a majority of the Galactic plane.  On the other hand, the survey suffers an increased incompleteness in regions of extreme extinction, such as the Galactic center where $n_{\rm H}$ exceeds $100\times 10^{20}$ cm$^{-2}$.  While a level of incompleteness is expected in such regions since the survey relies on optical follow-up observations, we demonstrate in the next section that a portion of this incompleteness is related to intrinsic limitations in the BSC from which CIZA targets are selected.
 
The redshift distribution of both the B2 subsample and the entire CIZA sample with detect fluxes above $3\times 10^{-12}$ erg cm$^{-2}$ s$^{-1}$ is shown in Figure 2.  The redshift distribution is qualitatively similar to that of the eBCS and REFLEX samples, both of which peak near a redshift of $z\sim0.08$.  



\begin{figure}[t]
\epsscale{1.2}
\plotone{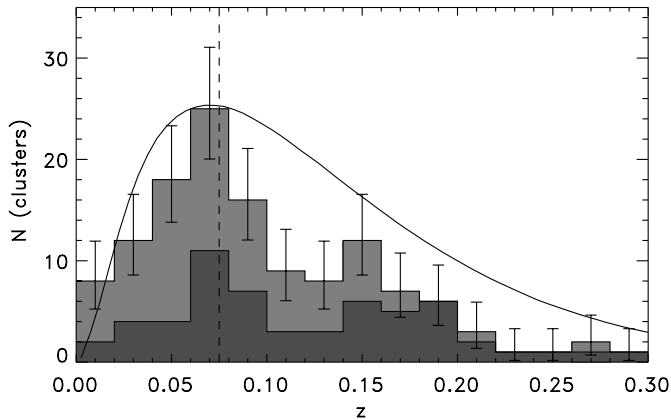}
\caption{Redshift distribution of the current CIZA sample.  The darker histogram includes the 60 B2 clusters with X-ray detect fluxes between 3 and $5\times 10^{-12}$ erg cm$^{-2}$ s$^{-1}$, while the lighter histogram includes all 133 CIZA clusters (B1+B2) with detect fluxes above $3\times 10^{-12}$ erg cm$^{-2}$ s$^{-1}$.  Four CIZA clusters have redshifts greater than 0.3 and fall outside the range of the figure.  The curved line shows the redshift distribution predicted by convolving the CIZA selection function with the X-ray luminosity function of the eBCS.  The dashed vertical line denotes z=0.075, below which we expect to be statistically complete.}
\end{figure}


\subsection{Survey Completeness}

We can estimate the completeness of the CIZA survey relative to the number of clusters we expect to detect in the BSC at low Galactic latitudes by using the cluster X-ray luminosity function measured outside the ZOA and the CIZA selection function.
The selection function, i.e. the solid angle over which a cluster with a given X-ray flux will be detected by the survey, can be accurately determined from the BSC's basic properties and the CIZA
selection criteria discussed in \S2.1.  The availability of a quantitative measure of the solid angle sampled
at a given brightness limit is one of the great advantages of X-ray
surveys over their optical counterparts.  

We begin by noting that the BSC has an intrinsic completeness limit of
17 source photons within the detect cell (Ebeling et al. 2001).  
Using the RASS exposure map \footnote{ Available in various formats from
http://www.xray.mpe.mpg.de/rosat/survey/rass-3/sup/} within the CIZA study
area we can translate this photon limit for secure detection into a
count rate selection function.  We next truncate this selection function to count rates above 0.05 counts s$^{-1}$, which is the intrinsic lower limit for detect rates included in the BSC (Voges et al. 1999).  Adopting an optically
thin, diffuse plasma with an average intra-cluster gas temperature of
4 keV as a model of the cluster X-ray emission and using the
Galactic $n_{\rm H}$ from Dickey \& Lockman (1990) in
the direction of each sampled point in our study area, as well as the
correct response function for the PSPC detector used in the RASS, we
convert this truncated count rate selection function into the unabsorbed BSC detect flux selection function shown in Figure 3.

\begin{figure}[t]
\epsscale{1.2}
\plotone{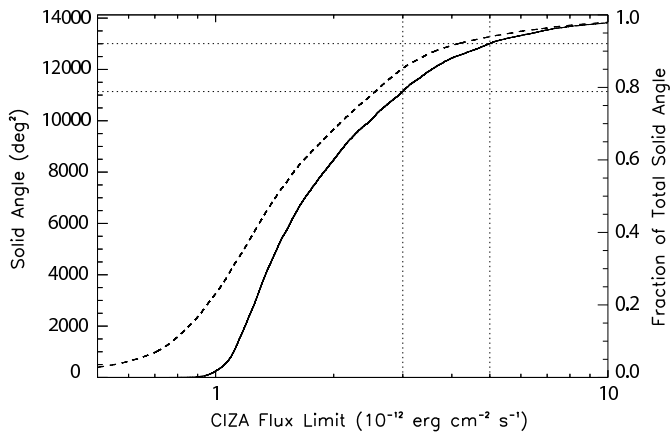}
\caption{Selection function of the CIZA survey, i.e., the fraction of the survey
solid angle covered at a given detect flux.  The dashed line shows the
selection function applicable to our survey if no count rate limit
were set; the solid curve represents the actual selection function
impacted by the intrinsic limit of 0.05 ct s$^{-1}$ to
the detect count rates included in the BSC. The dotted lines
mark the flux limits of the B1 and B2 subsamples.}
\end{figure}


\begin{figure*}[t]
\epsscale{1.1}
\plotone{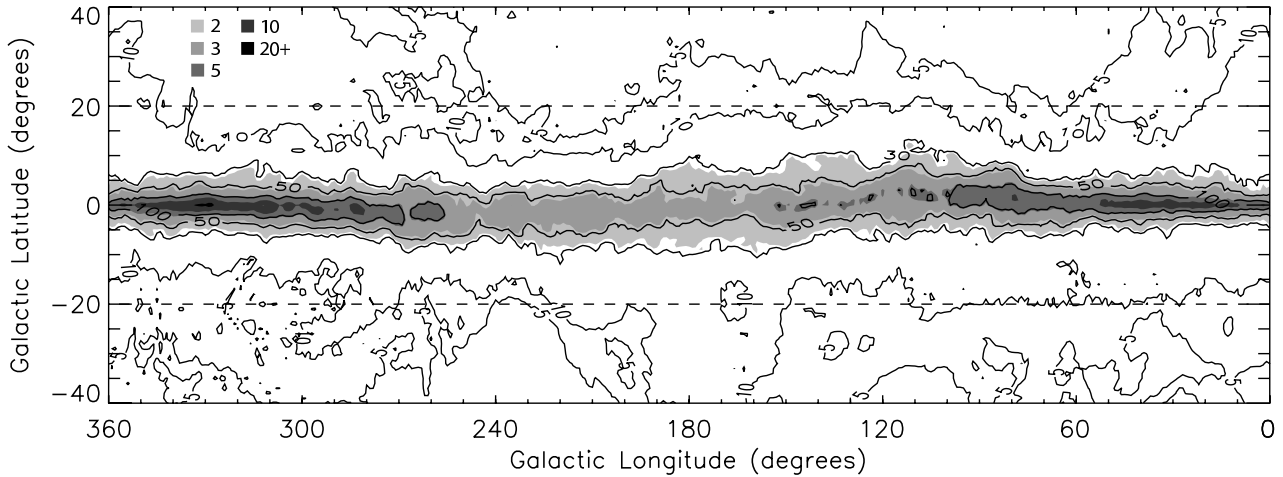}
\caption{Sensitivity map of the BSC catalog in the CIZA study area.  Solid lines are contours of Galactic $n_{\rm H}$ in units of $10^{20}$ cm$^{-2}$, while shaded regions are contours of limiting BSC detect flux for clusters behind the MW in units of $10^{-12}$ erg cm$^{-2}$ s$^{-1}$.  The relatively high flux limit toward the center of the Galaxy reflects the fact that only the brightest X-ray clusters will have observed count rates which exceed the minimum count rate limit intrinsic to the BSC.}
\end{figure*}

Predictions for the statistical completeness of each CIZA subsample can be calculated by convolving this selection function with the X-ray luminosity function of the eBCS sample (Ebeling et al. 2000).  This effectively allows us to calculate the number of clusters we expect to find in the BSC catalog at a certain redshift.  This prediction is shown in Figure 2 along with the redshift distribution of the current CIZA sample.  We find that down to the B2 flux limit, we are 54.0\% complete out to $z\sim0.3$ and 76.9\% complete out to $z\sim0.075$.  The number of clusters expected versus the number found in both the B1 and B2 subsamples is listed in Table 1.  

\begin{center}
\hspace*{-1in}
\begin{deluxetable}{cccccc}
\tablenum{1}
\tablewidth{0pc}
\tablecaption{Current completeness estimates for subsamples of the CIZA survey}
\tablecolumns{6}
\tablehead{\colhead{Sample} & \colhead{Flux}  & \colhead{BSC} & \colhead{Expected} & \colhead{Discovered}  & \colhead{Comp.\tablenotemark{$\beta$}} \\ \colhead{} & \colhead{Limit\tablenotemark{$\alpha$}} & \colhead{Targets} & \colhead{Clusters\tablenotemark{$\beta$}} & \colhead{Clusters\tablenotemark{$\beta$}} & \colhead{(\%)}}
\startdata
B1  &  5  & 481 & 101 (39) & 73 (41) & 72.3 (100.0)   \nl
B2  &  3  & 336 & 111 (26) & 60 (20) & 54.0 (76.9)  \nl
\enddata
\tablenotetext{$\alpha$}{Units of $10^{-12}$ erg cm$^{-2}$ s$^{-1}$}
\tablenotetext{$\beta$}{Values in parentheses are for clusters with $z < 0.075$}
\end{deluxetable}
\end{center}

\vspace*{-0.39in}

The prediction shown in Figure 2 takes into account an inherent limitation of the BSC at low Galactic latitudes, namely that it misses extragalactic sources in the central regions of the MW.  Since the BSC imposes a minimum count rate limit on sources included in the catalog, only the brightest X-ray clusters will have observed count rates (i.e. after attenuation through the plane) which exceed the minimum count rate limit in regions of extreme extinction.  This in effect means the flux limit of BSC sources seen through the MW varies with $n_{\rm H}$ along the Galactic plane, with the highest flux limits corresponding to regions with the highest $n_{\rm H}$ column densities.  In such areas of the plane we expect to miss clusters since relatively faint sources will be excluded from the BSC catalog from which we select CIZA targets.  We can identify the regions of the Galactic plane in which we expect to have an increased incompleteness by constructing a two dimensional map of the unabsorbed, detect flux limit of BSC sources over the entire CIZA study area.  This sensitivity map is shown in Figure 4; the solid lines are contours of Galactic $n_{\rm H}$, while the shaded contours represent the BSC detect flux limit for extragalactic sources.  In the very central regions of the plane, for example, only sources with X-ray fluxes greater than $20\times 10^{-12}$ erg cm$^{-2}$ s$^{-1}$ will have observed count rates that exceed the lower limit adopted in the BSC.  Roughly 15.2\% of the plane has a flux limit in excess of $3\times 10^{-12}$ erg cm$^{-2}$ s$^{-1}$, the nominal flux limit of the B2 subsample; in these regions we expect to have an increased incompleteness in the B2 subsample.  Therefore, while a level of incompleteness is expected in the CIZA sample at very low Galactic latitudes since the survey relies on optical follow-up observations, a significant portion of the survey's incompleteness is due to the fact that the BSC intrinsically misses faint extragalactic sources in the central regions of the MW.

\section{The B2 Sample}

In Table 2 we list the CIZA B2 sample. The
information provided in the various columns is as follows: column (1)
CIZA cluster name; column (2) alternative cluster name; column (3)
right ascension (J2000) of cluster X-ray centroid computed as
described in \S3.1; column (4) declination (J2000) of
cluster X-ray centroid computed as described in \S3.1;
column (5) Galactic longitude in degrees; column (6) Galactic latitude
in degrees; column (7) total cluster count rate in the \emph{ROSAT} broad
band in ct s$^{-1}$, computed as detailed in \S3.2;
column (8) statistical error of the total cluster count rate
($1\sigma$); column (9) equivalent neutral hydrogen column density
$n_{\rm H}$ in units of $10^{20}$ cm$^{-2}$ from Dickey \& Lockman
(1990); column (10) total cluster flux in units of $10^{-12}$ erg
cm$^{-2}$ s$^{-1}$, assuming a metallicity of 0.3 solar
and a cluster gas temperature consistent with the $L_{\rm X}-T$
relation of White, Jones \& Forman (1997); column (11) redshift;
column (12) total cluster X-ray luminosity in units of $10^{44}$
erg s$^{-1}$; column (13) redshift reference.  

\subsection{Cluster Coordinates}
 
The right ascension and declination of each cluster listed in Table 2 is determined from the centroid of the cluster's X-ray emission in the original RASS data.  The centroids are taken as the position of the X-ray peak within a $2\times 2$ $h_{\rm
50}^{-2}$ Mpc$^2$ box around each cluster (after convolving the RASS
count rate image with a Gaussian kernel of $\sigma=4$ arcmin width). These new positions differ
by up to 3 arcmin from the source positions listed in the
BSC. However, when translated into metric values, the offsets are much
less dramatic -- we find the metric positional differences to be on the order of 
50 $h_{\rm 50}^{-1}$ kpc. We visually inspect all images to ensure that
our cluster peak-finding procedure is not adversely affected by any
nearby bright point sources.



\subsection{Detect versus Total Cluster Fluxes} 

As previously mentioned the X-ray detect fluxes used for initial target selection
are based on the count rates listed in the BSC for
circular detection aperture of roughly 5 arcmin radius.
While these detect fluxes are accurate for point sources and distant clusters, they often underestimate the X-ray emission of nearby clusters due to their significant angular extent on the sky.  At the distance of the GA ($z\sim0.02$), for example, the standard BSC detect cell contains less than a quarter of the total X-ray flux from a cluster whose radial X-ray
surface brightness profile follows the canonical $\beta$ model
($r_{\rm c}=250$ kpc, $\beta=2/3$; Cavaliere \& Fusco-Femiano 1976).

Rather than extrapolating the BSC detect values to larger radii,
we obtain estimates of the total cluster count rates (and thus total fluxes and luminosities) directly from the RASS data.
We compute the total cluster count rates within circular apertures of
1.5 $h_{\rm 50}^{-1}$ Mpc radius around the improved X-ray
centroids. We then subtract the background count rate, as measured in an annulus from 2 to 3 $h_{\rm
50}^{-1}$ Mpc around the cluster center\footnote{We take the per-pixel
background count rate to be the mean of the distribution observed in
the background annulus, using iterative $3\sigma$ clipping to minimize
the impact of point sources.} to obtain net count rates for each cluster. 
Gaussian errors are computed and propagated based on
the photon statistics in the source and background regions.  Our total
cluster count rates are typically 1.29 times (median)
higher than the count rates listed in the BSC; this increases to 1.79 for clusters at $z<0.1$.  Figure 5 shows the ratio between the our recalculated total fluxes and the BSC detect fluxes versus redshift.  As expected the most nearby clusters show the largest discrepancy between the two values.  In the most extreme case, CIZAJ1614.3$-$6052 (A3627), our total count rate is more than 11 times higher than the BSC value, but still in good agreement with the value reported by B\"ohringer et al.\ (1996).

\section{Cosmography}

The compilation of the CIZA sample provides an opportunity to study the large-scale distribution, or cosmography, of the cluster population in the ZOA for the first time.  It has long been thought that the large-scale structures observed above and below the Galactic plane continue behind the MW (Tully et al. 1992).  We here examine the degree to which known superclusters extend into the ZOA and highlight newly discovered structures which have previously gone unnoticed.  Three regions have been of particular interest to the CIZA project due to their location at low Galactic latitudes: the GA region, the Perseus-Pisces (PP, Bernheimer 1932) concentration\footnote{This region has also been referred to as the Perseus-Pegasus concentration.}, and the Shapley supercluster (SSC, Shapley 1930).  The GA and SSC regions lie near to each other in projection toward the Galactic center, while the PP region is located on the opposite side of the Galactic plane (i.e. toward the Galactic anticenter).  The GA and the SSC are unique structures in the cluster distribution such that the former is purported to be the largest overdensity in the local universe, while the latter is the richest structure out of all the 220 identified superclusters out to $z = 0.12$ (Einasto et al. 1997), containing more than 4 times the number of rich clusters currently observed in the GA complex. 

\begin{figure}[t]
\epsscale{1.2}
\plotone{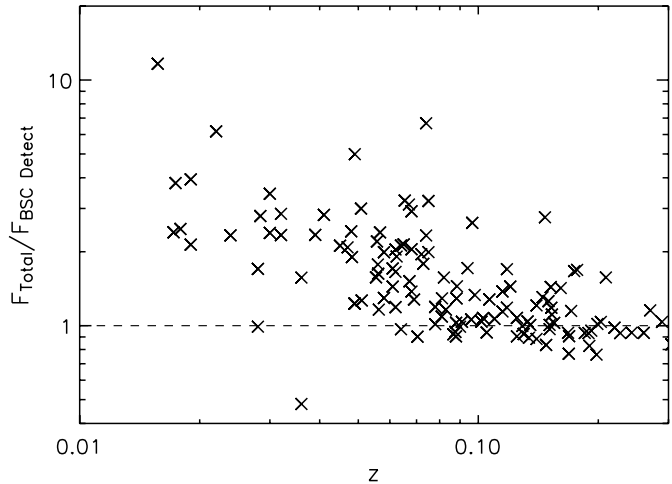}
\caption{Ratio of our recalculated total flux to the BSC detect flux as a function of redshift for the current CIZA sample.  The greatest discrepancy is seen at low redshifts, where the fixed aperture BSC detect fluxes underestimate the total X-ray emission of extended clusters.}
\end{figure}

In Figure 6 we show the distribution of CIZA clusters near the GA, PP, and SSC regions in supergalactic coordinates.  Also shown are eBCS and REFLEX clusters which have comparable X-ray fluxes.   We find that while newly discovered CIZA clusters extend all three regions into the ZOA, the addition of the B2 sample has led to a significant increase in the number of clusters \emph{behind} the GA.  Most notable among these are a set of 6 clusters which cross the ZOA near the SSC.  This association of clusters includes the well-known Triangulum Australis cluster (C1638\footnote{Hereafter we shorten Abell to 'A' and refer to CIZA clusters simply by their right ascension, i.e. CIZAJ1638.3-6421 becomes C1638.}), as well as C1410, C1514, C1518, C1614, C1652, and C1645.  One of these clusters, C1410, is near enough to the SSC to be considered a member of the supercluster.  The smooth transition from SSC members to CIZA clusters in this region suggests this string of clusters may trace an extension of the filament network in which the SSC is embedded.  The location of this set of clusters on the Galactic plane is highlighted in Figure 7.  We note that the string of clusters passes directly behind a region of extremely high extinction, which most likely explains the large separation between C1518 and C1614.  We suspect additional members of this large-scale filament remain hidden behind the very central regions of the Galactic plane, although their detection will have to await hard X-ray imaging, which is unaffected by extinction even at the lowest Galactic latitudes.



\begin{figure}[t]
\epsscale{1.2}
\plotone{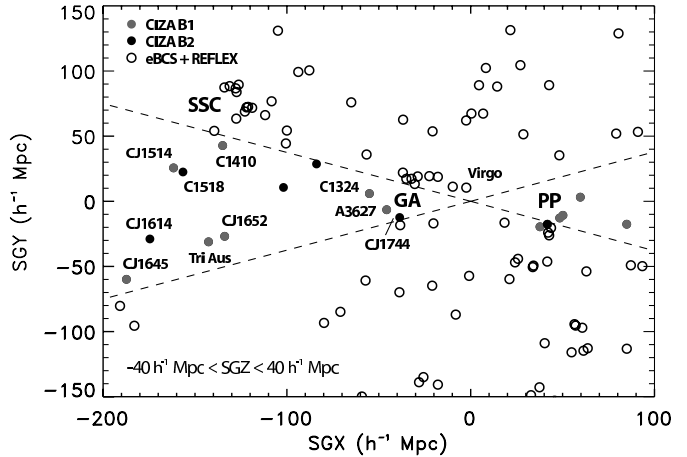}
\caption{Supergalactic projection of the CIZA cluster distribution near the GA, PP, and SSC superclusters, in a region $80 h^{-1}$ Mpc thick in SGZ.  Also shown are eBCS and REFLEX clusters with similar X-ray fluxes.  The dashed lines denote the boundary of the traditional ZOA.  Note the string of CIZA clusters which cross the ZOA near the SSC}
\end{figure}

Surprisingly, we find only one additional cluster in the GA region: C1744 at $b = -15.9$ and $z = 0.014$.  Although the survey is still in progress, we have thus far detected far fewer rich cluster systems in the GA complex than would be expected given the regions proposed mass.  Unless a set of highly obscured clusters are found in the next CIZA subsample, our preliminary findings suggest the GA cannot harbor the amount of mass that has traditionally been attributed to it.  Even with the discovery of rich clusters such as Norma (Abell 3627, Kraan-Korteweg et al. 1996), C1324 (Paper I), and C1744 near the Hydra-Centaurus region, there is almost an order of magnitude discrepancy between the mass concentration currently observed in the GA and the mass originally proposed by LB88.  Our findings of a less-massive GA and the detection of significant structures behind the complex supports studies which suggest the motion of nearby galaxies is due, in part, to a large-scale bulk flow which continues past the GA and is induced by more distant structures (e.g. Hudson 2005).   This would explain the reported lack of backside infall into the GA: galaxies on the far side of the GA are flowing into a yet more massive overdensity near the SSC region.  


\begin{figure}[t]
\epsscale{1.2}
\plotone{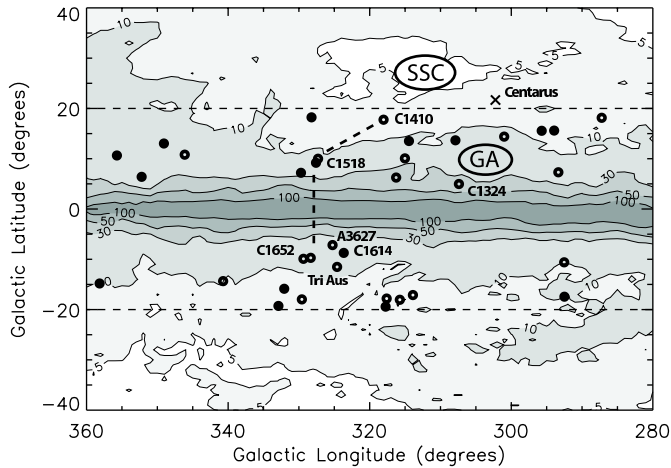}
\caption{Distribution of CIZA clusters along the Galactic plane near the SSC and GA regions.  The contours are the same as in Figure 1.  The filled circles are B2 clusters, while the open circles are B1 clusters.  The dashed line connects the string of clusters highlighted in Figure 6, which extend from the SSC to regions well inside the ZOA.}
\end{figure}

\begin{figure}[t]
\epsscale{1.2}
\plotone{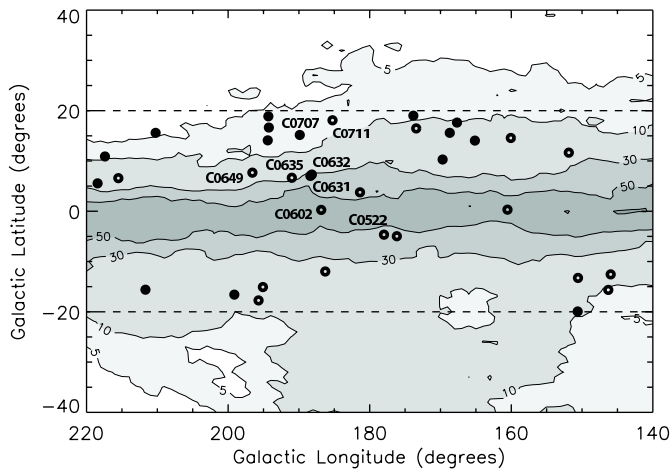}
\caption{Distribution of CIZA clusters along the Galactic plane near the Galactic anticenter.  The contours are the same as in Figure 1.  The filled circles are B2 clusters, while the open circles are B1 clusters.  The 8 clusters which make up the XXX supercluster are highlighted.}
\end{figure}

Finally, an additional structure of interest is a previously unknown association of 8 clusters that cross the ZOA near the Galactic anticenter.  Two of these clusters are from the newly released B2 sample.  The distribution on these clusters on the plane of the Galaxy is shown in Figure 8.  This structure is the first supercluster found to be completely hidden by the MW.  The association includes the C0522, C0602, C0631, C0632, C0635, C0649, C0707, and C0711 clusters, which have a median redshift of $z = 0.066$.  All but one of these clusters are newly discovered systems, with the exception being C0631, which is a redetection of the Zwicki cluster 1121.  This association is the most notable concentration of clusters in the current CIZA sample that is not correlated with a previously known supercluster region.

\section{Continuing Work \& Conclusions}

Although a handful of galaxy clusters have previously been found in the ZOA serendipitously, the CIZA survey is carrying out the first systematic search for clusters at low Galactic latitudes.  Thus far the survey has spectroscopically confirmed 205 galaxy clusters behind the Galactic plane, 60 of which are presented in this paper.  With the CIZA project proving effective in finding X-ray luminous clusters in the ZOA, the survey is providing the means to study the LSS behind the MW for the first time.  The current CIZA sample has already begun to shed light on previously hidden large-scale cluster associations and we may soon be able to settle the debate regarding the nature of the GA region.  Our preliminary findings suggest the GA contains far fewer rich cluster systems than would be expected given the regions proposed mass.  Having surveyed the entire GA complex, we find an order of magnitude discrepancy remains between the mass observed in the supercluster and that implied by the peculiar motion of nearby galaxies.  This inconsistency may be resolved by our detection a significant overdensity of clusters which appear near to the GA in projection, but are in fact four times as distant.  These include clusters between the GA and the SSC superclusters, as well as a large-scale association of clusters which may trace an extension of the SSC into the ZOA.  These findings support recent studies which suggest the peculiar velocity of nearby galaxies is only partly a result of infall into the GA, with the remaining motion due to a large-scale bulk flow induced by more distant overdensities in the ZOA.

To investigate these issues further we are currently extending the CIZA survey to fainter fluxes.  Our motivation is two-fold: (1) since the BSC detect fluxes are known to underestimate the total X-ray emission of low redshift clusters, it is possible that clusters in the nearby GA region have dropped below the flux limit of the B2 subsample, and have thus far been systematically overlooked, and (2) due to the bias in the BSC detect fluxes, the CIZA \emph{detect} flux limit is not equivalent to that of the eBCS and REFLEX samples covering the extragalactic sky since they are both flux limited down to a \emph{total} X-ray flux of $3\times10^{-12}$ erg cm$^{-2}$ s$^{-1}$.  To address both of these issues we are currently completing the CIZA survey down to a BSC detect flux of $2\times10^{-12}$ erg cm$^{-2}$ s$^{-1}$. In addition to containing intrinsically fainter clusters, we expect this B3 subsample to hold nearby, extended clusters whose total X-ray emission has been underestimated.  The B3 subsample currently contains 57 X-ray luminous clusters and we expect to complete the follow-up observation of our final cluster candidates shortly.  With the completion of this sample and ultimately the CIZA survey, we will be able to conclusively determine the extent of the GA region and combine the CIZA sample with the eBCS and REFLEX catalog to produce the first X-ray selected, all-sky cluster sample.

\acknowledgments
We thank the NASA Graduate Student Research Program for funding this research and the University of Hawaii's Time Allocation Committee for their generous support of the CIZA survey.

\clearpage
\begin{deluxetable}{llccrrrrrrcrr}
\tablenum{2}
\tablecaption{Catalog of the 60 CIZA clusters with BSC X-ray detect fluxes between 3 and $5\times10^{-12}$ erg cm$^{-2}$ s$^{-1}$ (0.1--2.4 keV)}
\tablecolumns{13}
\tablewidth{0pc} 
\tablehead{
\colhead{CIZA Name} &
\colhead{Other Name} &
\colhead{R.A.$^a$}  & 
\colhead{Dec.$^a$}  &
\colhead{$l$ (deg)}  &
\colhead{$b$ (deg)}  &
\colhead{$cr_{\rm X}$$^b$}  &
\colhead{$\Delta cr_{\rm X}$$^b$}  &
\colhead{$n_{\rm H}$$^c$}  &
\colhead{$f_{\rm X}$$^d$}  &
\colhead{$z$}  &
\colhead{$L_{\rm X}$$^e$}  &
\colhead{$z$ Ref.}} 
\tablecomments{Count rates, fluxes, and luminosities are total values within a radius of 1.5 $h_{\rm 50}^{-1}$ Mpc; 
               $^a$ J2000, 
               $^b$ in units of ct s$^{-1}$, 
               $^c$ in units of $10^{20}$ cm$^{-2}$, 
               $^d$ in units of $10^{-12}$ erg cm$^{-2}$ s$^{-1}$ (0.1--2.4 keV), 
               $^e$ in units of $10^{44}$ erg s$^{-1}$ (0.1--2.4 keV)\\ 
               \\
               References: (1) this work; (2) Struble \& Rood (1991); (3) Crawford et al. (1995); 
               (4) Hewitt \& Burbidge (1991); (5) Allen et al. (1992)}
\startdata 
CIZAJ0142.9$+$4438  &			&  01 42 59.9 & $+44$ 38 33  & 132.500   &  -17.269  &  0.2128  &   0.0455   &    9.12  &    4.75  &   0.3410  &   22.49  &  (1)  \\
CIZAJ0157.0$+$4120 & A276	      	&  01 57 07.8 & $+41$ 20 24  & 135.976   &  -19.862  &  0.2455  &   0.0651   &    8.05  &    5.31  &   0.0811  &    1.50  &  (2)  \\
CIZAJ0301.8$+$3550  & A407	      	&  03 01 47.7 & $+35$ 49 57  & 150.611   &  -19.937  &  0.3251  &   0.0557   &   11.56  &    7.84  &   0.0470  &    0.74  &  (2)  \\
CIZAJ0527.2$+$0412  &	   	      	&  05 27 13.7 & $+04$ 13 08  & 199.119   &  -16.576  &  0.1800  &   0.0296   &   13.48  &    4.54  &   0.1510  &    4.39  &  (1)  \\
CIZAJ0553.4$-$0557  &	      	    	&  05 53 27.0 & $-05$ 58 04  & 211.683   &  -15.604  &  0.1526  &   0.0262   &   19.91  &    4.45  &   0.1510  &    4.31  &  (1)  \\
CIZAJ0604.6$+$4256  &	      	    	&  06 04 39.6 & $+42$ 57 10  & 169.688   &   10.293  &  0.2176  &   0.0403   &   21.65  &    6.63  &   0.1180  &    3.92  &  (1)  \\
CIZAJ0612.5$+$4836  & A553	    	&  06 12 36.8 & $+48$ 36 07  & 165.148   &   14.053  &  0.3749  &   0.0613   &   15.08  &    9.86  &   0.0670  &    1.90  &  (2)  \\
CIZAJ0616.3$-$2156  &	   	      	&  06 16 22.9 & $-21$ 56 54  & 229.240   &  -17.230  &  0.1664  &   0.0267   &   11.12  &    3.94  &   0.1710  &    4.88  &  (1)  \\
CIZAJ0629.1$+$4606  &	      	    	&  06 29 09.8 & $+46$ 06 04  & 168.695   &   15.586  &  0.1525  &   0.0369   &   12.38  &    3.74  &   0.1290  &    2.65  &  (1)  \\
CIZAJ0632.0$+$2519  &	      	    	&  06 32 02.0 & $+25$ 19 37  & 188.162   &    7.298  &  0.1736  &   0.0484   &   33.84  &    6.83  &   0.0750  &    1.65  &  (1)  \\
CIZAJ0638.0$+$4747  & ZwCl\,0634.1+4750	&  06 38 07.5 & $+47$ 47 28  & 167.664   &   17.640  &  0.2921  &   0.0409   &   10.41  &    6.77  &   0.1740  &    8.63  &  (3)  \\
CIZAJ0657.0$+$4246  &	   		&  06 57 00.7 & $+42$ 46 49  & 173.830   &   19.001  &  0.1284  &   0.0338   &    8.97  &    2.86  &   0.1250  &    1.91  &  (1)  \\
CIZAJ0707.0$+$2706  &	   		&  07 07 02.6 & $+27$ 06 23  & 189.881   &   15.169  &  0.3229  &   0.0669   &    7.41  &    6.83  &   0.0620  &    1.13  &  (1)  \\
CIZAJ0710.4$+$2240  &	   		&  07 10 21.4 & $+22$ 40 24  & 194.393   &   14.092  &  0.1720  &   0.0277   &    6.73  &    3.55  &   0.2890  &   12.29  &  (1)  \\
CIZAJ0720.3$+$2349  &	   		&  07 20 18.0 & $+23$ 48 57  & 194.245   &   16.642  &  0.1862  &   0.0328   &    6.37  &    3.79  &   0.1680  &    4.53  &  (1)  \\
CIZAJ0721.2$-$0220 &	   		&  07 21 13.4 & $-02$ 20 33  & 218.433   &    5.544  &  0.1945  &   0.0885   &   17.16  &    5.42  &   0.0360  &    0.30  &  (1)  \\
CIZAJ0729.4$+$2436  &	   		&  07 29 28.8 & $+24$ 36 21  & 194.315   &   18.878  &  0.2490  &   0.0359   &    6.71  &    5.14  &   0.1610  &    5.63  &  (1)  \\
CIZAJ0738.2$+$0102  &	   		&  07 38 16.6 & $+01$ 01 51  & 217.401   &   10.883  &  0.1875  &   0.0320   &    8.47  &    4.11  &   0.1890  &    6.19  &  (1)  \\
CIZAJ0742.6$+$0922  & A592	   	&  07 42 38.6 & $+09$ 22 09  & 210.242   &   15.593  &  0.3692  &   0.0748   &    3.57  &    6.37  &   0.0624  &    1.07  &  (2)  \\
CIZAJ0745.1$-$5404  & ESO 163-IG015	&  07 45 09.6 & $-54$ 04 45  & 266.844   &  -14.359  &  0.4452  &   0.0389   &   12.29  &   10.89  &   0.0740  &    2.55  &  (1)  \\
CIZAJ0757.9$-$2157  &	   		&  07 57 58.8 & $-21$ 57 44  & 239.991   &    3.771  &  0.5441  &   0.1131   &   28.43  &   19.22  &   0.0490  &    1.98  &  (1)  \\
CIZAJ0802.1$-$1926  &	   		&  08 02 09.6 & $-19$ 26 02  & 238.336   &    5.923  &  0.1128  &   0.0425   &   16.67  &    3.08  &   0.1400  &    2.57  &  (1)  \\
CIZAJ0809.9$-$0258  &	   		&  08 09 55.7 & $-02$ 58 11  & 224.879   &   15.953  &  0.2380  &   0.0413   &    4.53  &    4.45  &   0.2700  &   13.44  &  (1)  \\
CIZAJ0815.4$-$0308  & 3C196.1	   	&  08 15 28.6 & $-03$ 08 31  & 225.746   &   17.074  &  0.1503  &   0.0390   &    6.51  &    3.08  &   0.1980  &    5.10  &  (4)  \\
CIZAJ0943.3$-$7619  & PKS 0943-76	&  09 43 25.0 & $-76$ 19 57  & 292.495   &  -17.438  &  0.2132  &   0.0336   &   10.56  &    4.97  &   0.1990  &    8.26  &  (1)  \\
CIZAJ1201.0$-$4623  &	   		&  12 01 03.4 & $-46$ 23 27  & 293.948   &   15.600  &  0.1908  &   0.0442   &    6.86  &    3.95  &   0.1180  &    2.35  &  (1)  \\
CIZAJ1210.6$-$4644  & AS689	   	&  12 10 44.2 & $-46$ 44 19  & 295.712   &   15.562  &  0.4572  &   0.1158   &    7.20  &    9.56  &   0.0320  &    0.42  &  (1)  \\
CIZAJ1320.7$-$4855  &	   		&  13 20 47.0 & $-48$ 55 49  & 307.885   &   13.659  &  0.1174  &   0.0309   &   12.84  &    2.90  &   0.2270  &    6.30  &  (1)  \\
CIZAJ1358.7$-$4750  &	   		&  13 58 56.9 & $-47$ 50 19  & 314.493   &   13.512  &  0.9306  &   0.1088   &   11.90  &   22.50  &   0.0740  &    5.25  &  (1)  \\
CIZAJ1456.2$-$3826  &	   		&  14 56 12.2 & $-38$ 26 25  & 328.215   &   18.230  &  0.2399  &   0.0528   &    6.41  &    4.88  &   0.1150  &    2.75  &  (1)  \\
CIZAJ1518.4$-$4631  &	   		&  15 18 22.8 & $-46$ 32 26  & 327.562   &    9.185  &  0.2449  &   0.0687   &   18.85  &    7.08  &   0.0560  &    0.95  &  (1)  \\
CIZAJ1535.1$-$4659  &	   		&  15 35 09.1 & $-46$ 58 45  & 329.709   &    7.203  &  0.0603  &   0.1035   &   27.24  &    2.11  &   0.0360  &    0.12  &  (1)  \\
CIZAJ1614.1$-$6306  &	   		&  16 14 07.9 & $-63$ 07 51  & 323.646   &   -8.734  &  0.2307  &   0.0913   &   16.61  &    6.32  &   0.0620  &    1.04  &  (1)  \\
CIZAJ1627.8$-$2952  &	   		&  16 27 58.8 & $-29$ 52 51  & 349.082   &   12.999  &  0.1119  &   0.0671   &   15.25  &    2.97  &   0.0640  &    0.53  &  (1)  \\
CIZAJ1655.0$-$2624  &	   		&  16 55 01.7 & $-26$ 25 29  & 355.697   &   10.645  &  0.2522  &   0.0613   &   16.12  &    6.78  &   0.0940  &    2.56  &  (1)  \\
CIZAJ1700.7$-$3144  &	   		&  17 00 53.5 & $-31$ 44 30  & 352.229   &    6.384  &  0.1007  &   0.0479   &   18.02  &    2.84  &   0.1340  &    2.18  &  (1)  \\
CIZAJ1705.8$-$7422  &	   		&  17 05 58.1 & $-74$ 22 31  & 317.758   &  -19.431  &  0.1457  &   0.0533   &    7.19  &    3.06  &   0.1900  &    4.68  &  (1)  \\
CIZAJ1744.9$-$6044 & NGC 6407	   	&  17 44 57.6 & $-60$ 44 06  & 332.074   &  -15.864  &  0.1785  &   0.4814   &    7.89  &    3.30  &   0.0140  &    0.03  &  (1)  \\
CIZAJ1809.0$-$0414  &	  		&  18 09 04.3 & $-04$ 14 39  & 24.2636   &    7.435  &  0.1171  &   0.0286   &   27.92  &    4.01  &   0.3050  &   15.35  &  (1)  \\
CIZAJ1813.3$-$6126  & PMN J1813-6126	&  18 13 18.7 & $-61$ 27 27  & 332.879   &  -19.278  &  0.4061  &   0.0850   &    7.97  &    8.75  &   0.1470  &    7.98  &  (1)  \\
CIZAJ1824.1$+$3029 & NPM1G +30.0453	&  18 24 07.7 & $+30$ 29 33  &  58.257   &   18.817  &  0.2996  &   0.0406   &    8.61  &    6.60  &   0.0720  &    1.47  &  (1)  \\
CIZAJ1839.8$-$2108  &	    		&  18 39 53.5 & $-21$ 08 38  &  12.747   &   -7.047  &  0.1868  &   0.0658   &   20.42  &    5.59  &   0.0680  &    1.11  &  (1)  \\
CIZAJ1844.5$-$3724  & CGMW 4-2867	&  18 44 31.7 & $-37$ 23 33  & 358.159   &  -14.808  &  0.1875  &   0.0567   &    9.74  &    4.27  &   0.2030  &    7.39  &  (1)  \\
CIZAJ1904.2$+$3627  & MG2 J190411+3627	&  19 04 13.2 & $+36$ 26 48  &  67.295   &   13.355  &  0.1621  &   0.0343   &    9.05  &    3.63  &   0.0780  &    0.95  &  (1)  \\
CIZAJ1910.1$-$2238  & CGMW 3-3366	&  19 10 08.9 & $-22$ 39 04  &  14.394   &  -14.039  &  0.2135  &   0.0860   &   12.64  &    5.29  &   0.0563  &    0.72  &  (1)  \\
CIZAJ1916.0$+$3525  &	   		&  19 16 06.5 & $+35$ 25 20  &  67.381   &   10.741  &  0.2186  &   0.0263   &    9.87  &    4.99  &   0.2090  &    9.13  &  (1)  \\
CIZAJ1917.6$-$1315  & TXS 1914-133	&  19 17 35.8 & $-13$ 15 20  &  23.911   &  -11.725  &  0.2060  &   0.0404   &   14.67  &    5.33  &   0.1770  &    7.04  &  (1)  \\
CIZAJ1918.5$-$0842  & CGMW 3-3640	&  19 18 31.4 & $-08$ 42 29  &  28.180   &   -9.949  &  0.1401  &   0.0493   &   17.16  &    3.88  &   0.0900  &    1.35  &  (1)  \\
CIZAJ1925.3$+$3705  &	   		&  19 25 22.3 & $+37$ 05 50  &  69.746   &    9.806  &  0.1693  &   0.0236   &   13.72  &    4.27  &   0.3140  &   17.28  &  (1)  \\
CIZAJ1926.1$+$4832  &	   		&  19 26 10.8 & $+48$ 33 00  &  80.373   &   14.644  &  0.3188  &   0.0318   &    6.93  &    6.62  &   0.0980  &    2.72  &  (1)  \\
CIZAJ1947.6$-$0541  & CGMW 3-4849	&  19 47 39.4 & $-05$ 42 15  &  34.221   &  -15.093  &  0.1603  &   0.1226   &   10.97  &    3.74  &   0.0280  &    0.13  &  (1)  \\
CIZAJ1948.2$+$5114  &	   		&  19 48 19.2 & $+51$ 13 51  &  84.466   &   12.628  &  0.1717  &   0.0189   &   10.49  &    3.99  &   0.1850  &    5.77  &  (1)  \\
CIZAJ1957.2$+$5751 & ZwCl\,1956.0+5746	&  19 57 13.9 & $+57$ 51 13  &  91.107   &   14.596  &  0.2217  &   0.0251   &   11.64  &    5.34  &   0.0884  &    1.79  &  (5)  \\
CIZAJ2015.3$+$5609  &	   		&  20 15 18.7 & $+56$ 09 34  &  90.867   &   11.618  &  0.1648  &   0.0264   &   20.16  &    4.88  &   0.0820  &    1.41  &  (1)  \\
CIZAJ2106.2$+$3426  &	   		&  21 06 16.3 & $+34$ 26 25  &  78.996   &   -8.595  &  0.1534  &   0.0376   &   15.32  &    4.07  &   0.0866  &    1.31  &  (1)  \\
CIZAJ2242.8$+$5301  &	   		&  22 42 53.0 & $+53$ 01 06  & 104.191   &   -5.111  &  0.1140  &   0.0240   &   33.57  &    4.38  &   0.1921  &    6.80  &  (1)  \\
CIZAJ2302.6$+$7136  &	   		&  23 02 45.1 & $+71$ 37 15  & 114.510   &   10.549  &  0.1696  &   0.0243   &   27.43  &    5.81  &   0.1450  &    5.18  &  (1)  \\
CIZAJ2319.7$+$4251 & NGC 7618	   	&  23 19 47.5 & $+42$ 51 18  & 105.577   &  -16.907  &  0.7159  &   0.1628   &   11.76  &   17.25  &   0.0173  &    0.22  &  (1)  \\
CIZAJ2320.2$+$4146  & RGB J2320+417	&  23 20 14.4 & $+41$ 46 44  & 105.249   &  -17.940  &  0.2030  &   0.0341   &    9.78  &    4.63  &   0.1400  &    3.89  &  (1)  \\
CIZAJ2330.6$+$4556  &	   		&  23 30 39.8 & $+45$ 56 15  & 108.564   &  -14.669  &  0.1460  &   0.0259   &   12.36  &    3.56  &   0.2425  &    8.76  &  (1)  \\
\enddata 
  \end{deluxetable}
\clearpage

\end{document}